\documentstyle[12pt,aaspp4]{article}

\newcommand {\be} {\begin{equation}}
\newcommand {\ee} {\end{equation}}

\def\refitem{\par\parskip 0pt\noindent\hangindent 20pt}

\begin{document}

\title{On Pair Content and Variability of Sub-Parsec Jets in Quasars}
\author{M.~Sikora$^{1}$ and G.~Madejski$^{2,3}$}
\affil
{$^1$ Nicolaus Copernicus Astronomical Center, Bartycka 18, 00-716
Warsaw, Poland\\
$^2$ Laboratory for High Energy Astrophysics, NASA/GSFC, Greenbelt, MD
20771, USA\\
$^3$ Dept. of Astronomy, University of Maryland, College Park}

\vskip 2 cm

\centerline {\bf Accepted for publication in the Astrophysical Journal}

\vskip 2 cm

\begin{abstract}

X--ray observations of blazars associated with the 
OVV (Optically Violently Variable) quasars put strong 
constraints on the e$^+$e$^-$ pair
content of radio-loud quasar jets.  From those observations, we infer that
jets in quasars contain many more e$^+$e$^-$ pairs 
than protons, but dynamically are still dominated by protons.  
In particular, we show that pure e$^+$e$^-$ jet models can be 
excluded, as they overpredict soft X--ray radiation; likewise, 
pure proton-electron jets can be excluded, as they predict too
weak nonthermal X--ray radiation.  An intermediate case is viable.  
We demonstrate that jets which are initially proton-electron
(``proto-jets'') can be 
pair-loaded via interaction with 100 -- 300 keV photons produced in hot 
accretion disc coronae, likely to exist in active galactic nuclei in
general. If the coronal radiation is powered by magnetic flares,
the pair loading is expected to be non-uniform and non-axisymmetric.
Together with radiation drag, this leads to velocity and density 
perturbations in a jet and formation of shocks, where the pairs are
accelerated.  Such a scenario can 
explain rapid (time scale of $\sim$ a day) variability observed in
OVV quasars.

\end{abstract}


\section{INTRODUCTION}

One of the basic unresolved questions regarding the nature of jets in 
radio loud quasars
is that of their composition: are they made from protons  and electrons, 
or electron-positron pairs, or from a mixture of both? Arguments in 
favor of proton-electron jets in quasars have been recently advanced by 
Celotti \& Fabian (1993). Using synchrotron self-Compton constraints from 
radio-core observations and information about energetics of jets 
from radio-lobe studies, those authors showed that in the case of pure
electron-positron jets the required number 
of e$^+$e$^-$ pairs is too high to be delivered from the central engine. 
The limit is imposed by the annihilation process 
(Guilbert, Fabian, \& Rees 1983). 

On the other hand, the recently discovered circular polarization in
the radio cores of the $\gamma$--ray bright OVV quasar 3C 279 and several 
other objects and its interpretation via the 
``Faraday conversion'' process suggests that the jet plasma is
dominated by e$^+$e$^-$ pairs (Wardle et al. 1998; Wardle \& Homan 1999). 
The fact that jets are likely to be pair dominated has also been inferred
from synchrotron self-Compton analyses of compact radio components in 
radio galaxy M87 (Reynolds et al. 1996), and in quasar 3C 279 
(Hirotani et al. 1999). 
In this paper we derive constraints imposed on the pair content 
of quasar jets by X--ray observations of OVV quasars, i.e. those radio loud 
quasars which are observed at very small angles to the jet axis, and often 
detected in the MeV - GeV $\gamma$--ray regime.  Our results 
suggest that the pair content of quasar jets is high, but that dynamically 
the jets are dominated by protons. 

The question of composition of the jet plasma is closely related 
to that of the formation of the jet.  Jets can be launched as 
outflows dominated by Poynting flux, generated in the force-free 
magnetosphere of the black hole, or as hydromagnetic winds 
driven centrifugally from an accretion disc (see, e.g., a review 
by Lovelace et al. 1999).  Electromagnetically dominated outflows are 
converted to pair dominated jets (Romanova \& Lovelace 1997; 
Levinson 1998), whilst hydromagnetic winds give rise to 
proton-electron dominated jets (Blandford \& Payne 1982). 
While the pair-dominated jets are predicted to be relativistic, the 
proton-electron jets can be either relativistic or non-relativistic, 
depending on whether the magnetic forces dominate over gravity in the 
accretion disc corona (Meier et al. 1997).  In particular, relativistic
hydromagnetic jets can  be launched deeply in the ergosphere
of fast-rotating black holes (Koide et al. 1999). If this is the case,
and the surrounding accretion disc corona is as hot as inferred from
the spectra of Seyfert galaxies (see, e.g., Zdziarski et al. 1994; Matt 1999), 
the proton-electron jets can be pair-loaded via the interactions with the 
coronal hard X--rays / soft $\gamma$--rays.  We demonstrate that 
this process is efficient enough to provide the number of pairs
required to account for the observed X--ray spectra of OVV quasars.  
Furthermore, the rapid X--ray variability of Seyferts (Green et al. 
1993; Hayashida et al. 1998)  indicates that the corona is likely 
to have dynamical character (as would be if it is powered by magnetic 
flares), and thus the hydromagnetic outflows are expected to be loaded
by pairs nonuniformly and non-axisymmetrically. This, together with 
the radiation drag imposed by the coronal and disc radiation fields 
on pairs, can lead to a modulation of the velocity and density of 
the plasma in a jet, and therefore to production of shocks and 
acceleration of  particles responsible for the variable nonthermal 
radiation.

Our paper is organized as follows:
In \S 2 we  demonstrate
that pure pair jets overpredict the soft X--ray flux; in \S 3 we show
that jets which are dynamically dominated by protons must be heavily 
pair-loaded in order to provide the observed nonthermal X--ray radiation;
and in \S 4 we show that hydromagnetic outflows can be pair-loaded due to 
interaction with hard X--rays / soft $\gamma$--rays from the hot accretion disc 
corona. Our results are summarized in \S 5.   

\section{ELECTRON-POSITRON JETS}

\subsection{Cold pairs}

If the relativistic jet is dynamically dominated by cold 
e$^+$e$^-$ pairs, then the external UV photons will be Comptonized 
by those pairs and thus boosted in frequency by a square of a bulk Lorentz 
factor $\Gamma$ and collimated along the jet axis.  In this case, in 
addition to the nonthermal radiation from the jet -- which results in the
phenomenon called blazar -- the observer located at 
$\theta_{obs} \le 1/\Gamma$ should see a soft X--ray ``bump''
superimposed on the continuum in OVV quasars.  Such spectral feature, 
predicted theoretically by Begelman \& Sikora (1987), has not been 
observed, and this fact can be used to derive an upper limit for a 
pair content of quasar jets (Sikora et al. 1997).

Luminosity of the soft X--ray bump produced by the above ``bulk-Compton'' (BC) 
process is 
\be L_{BC} \simeq {\cal A} \int_{r_0} {1 - e^{-\tau_j} \over \tau_j} 
{\left \vert dE_e \over dt \right \vert} n_e \, dV , \label{Lbc} \ee
where 
\be {\left \vert dE_e \over dt \right \vert} = 
m_e c^2 {\left \vert d\Gamma \over dt \right \vert}=
{4 \over 3} c \sigma_T u_{diff} \Gamma^2
\, , \label{dEdt} \ee
\be u_{diff}= {\xi L_d \over 4 \pi r^2 c} \, , \label{udiff} \ee
${\cal A}$ is the beaming amplification factor, $L_d$ is the luminosity of an 
accretion disc, $\xi$ is the fraction of the
accretion disk which at the given distance $r$ is isotropized due to
rescattering or reprocessing,
$n_e$ is the density of electrons plus positrons,
 $dV=\pi a^2 dr$, $\tau_j = n_e a \sigma_T$,  
$a$ is the cross-section radius of a jet, and $r_0$ is the 
distance at which the jet is fully formed (accelerated and collimated).
 
Assuming that jet is conical and that at $r > r_0$ the electron/positron 
number flux is conserved (no pair production), we have
$n_e \propto 1/r^2$, and $\tau_j = r_1/r$, where  by $r_1$ we denote 
the distance
at which $\tau_j=1$. Then, for jets with a half-angle $\theta_j \equiv a/r
\le 1/\Gamma$, Eqs. (\ref{Lbc})-(\ref{udiff}) give
\be L_{BC} \simeq {1 \over 3} (\Gamma \theta_j) ( \xi L_d) \Gamma^3 
\cases {\ln {r_1 \over r_0} +1 & if $\tau_j (r_0) > 1 $ \cr
{r_1 \over r_0} & if $\tau_j (r_0) < 1$ \cr} \, , \label{Lbc2} \ee 
where we used ${\cal A} = \Gamma^2$.
The value of $r_1$ depends on the electron/positron number flux, $dN_e/dt$, 
which -- for energy flux in a jet $L_j$ dominated by kinetic energy flux
of cold pairs -- is equal to $L_j/m_ec^2\Gamma$. Noting that
$dN_e/dt  \simeq n_e c \pi a^2$, we obtain
\be r_1 = {1 \over \sigma_T n_e \theta_j}
\simeq {\sigma_T L_j \over \pi m_e c^3 \theta_j \Gamma}
\simeq 10^{17} \,{L_{j,46} \over \theta_j \Gamma} \, {\rm cm} \, , 
\label{r_T} \ee
If $r_0 > r_1$ then $r_1$ should be 
treated only as a formal parameter which provides normalization of
$\tau_j$. However, noting the very large value of $r_1$ one can expect that
$r_0 < r_1$ and, therefore, the predicted Bulk-Compton luminosity is
\be L_{BC} > 3 \times 10^{47}  (\Gamma \theta_j) ( \xi L_d)_{45} (\Gamma/10)^3
(\ln (r_1/r_0) + 1)
\, {\rm erg \, s}^{-1}\, . \label{Lbc3} \ee
The BC spectral component peaks at $h\nu \sim \Gamma^2 h\nu_{UV}
\simeq (\Gamma/10)^2$ keV, where typical luminosities observed  in 
OVV quasars are
$\sim 10^{46}$ erg s$^{-1}$ (Sambruna 1997), while the spectra are
consistent with simple power laws (cf. Kubo et al. 1998). Thus, one 
can conclude that {\sl pure electron-positron jet models can be excluded 
as overpredicting soft X--ray radiation of OVV quasars}. 

\subsection{Relativistic ``thermal'' pairs}

It is of course possible that because of inefficient cooling of 
electrons/positrons below a given energy, the multiple reacceleration
process balances adiabatic losses in the conically diverging jet, 
and the pairs, once accelerated, remain relativistic forever.
If the relativistic electrons are narrowly distributed
around some $\bar\gamma$, then taking into account that
$\vert d E_e/dt \vert \propto \Gamma^2 \bar\gamma^2$,
$n_e \propto L_j/\bar\gamma$, and 
$r_1 \simeq r_1 (\bar \gamma =1)/\bar \gamma$, and assuming $r_0 > r_1$,
the bulk-Compton luminosity
would be
\be L_{BC} = 3 \times 10^{47}  (\Gamma \theta_j) ( \xi L_d)_{45} (\Gamma/10)^3
(r_1/r_0) \bar \gamma \simeq
3 \times 10^{48} {( \xi L_d)_{45}(\Gamma/10)^3 L_{j,46} 
\over (r_0/10^{16} {\rm cm})}
\, {\rm erg \, s}^{-1}\, , \label{Lbc4} \ee 
and would peak at $\sim h\nu \simeq \Gamma^2 \bar\gamma^2 \nu_{UV} \simeq
\bar\gamma^2 (\Gamma/10)^2$~keV.  No such ``bumps'' have 
been detected at keV energies. Alternatively, one can speculate that the 
X--ray spectra of OVV quasars 
consist of superposed multiple ``thermal'' peaks produced
over several decades of distance (cf. Sikora et al. 1997).  In this case
it may be possible to match the observed  X--ray spectral 
slopes, but nonetheless, the model predicts too large luminosity.

\section{PROTON-ELECTRON JETS}

The other extreme would have no e$^+$e$^-$ pairs in a jet.  
For a given energy flux in the jet, $L_j$, which now is proportional
to $n_p m_p$, the number of electrons in a jet is $m_e /m_p$ times 
smaller than the number of electrons plus positrons in the jet made 
from cold pairs.  Thus, noting that $r_1(n_e=n_p)= (m_e/m_p) r_1(n_p=0) \simeq
0.5 \times 10^{14}$ cm, one can find that the proton-electron jets do 
not overproduce the soft X--ray luminosities observed in OVV quasars, 
provided that $r_0 \ge 15 \, r_1 \simeq 10^{15}$ cm.  

However, such pure proton-electron jets are relatively inefficient in 
producing the nonthermal radiation, and this is for the same reason 
as above -- low number of electrons.  This is apparent from a study of 
the low energy tails of the nonthermal radiation components, where the 
requirement for the number of electrons is largest.  In the case of 
synchrotron radiation, such tails are not observed because they 
are self-absorbed, and the only spectral band where the presence of 
the lower energy relativistic electrons can be evident are the 
soft and mid-energy X--rays, 0.1 -- 20 keV.  

It has been shown in many papers (see, e.g., Sikora, Begelman, \& Rees
1994) that the $\gamma$--ray spectra observed by the EGRET instrument in 
OVV quasars are likely to 
be produced by Comptonization of external diffuse radiation field, 
via the so-called external radiation Compton (ERC) process.  Can this 
process be also responsible for the X--ray spectra of OVV quasars?  In the 
one zone model and for the narrow spectral distribution of the soft 
radiation field, the  power law X--ray spectra are produced by electrons
with energy distribution obeying $n_{\gamma}' = C_n \gamma^{-s}$, 
where $s=2 \alpha_X +1$ (note that throughout this paper, all
quantities except for $\gamma$ are primed if measured in the frame
co-moving with the jet).
Assuming that at a  distance $r_{fl}$, where the 
blazar phenomenon is produced, all available electrons are accelerated 
and that energy flux in the  jet is dominated by cold protons, i.e., that 
\be n_e' = \int_{\gamma_{min}}n_{\gamma}'d\gamma = 
n_p' \simeq  {L_j \over   m_p c^3 \Gamma^2 \pi a^2} \, , 
\label{nre} \ee  
we obtain
\be C_n = 
{(s-1)\, \gamma_{min}^{s-1} \, L_j \over m_p c^3 \pi\, a^2 \, \Gamma^2} 
\, . \label{Cn} \ee
The ERC luminosity is then given by 
$$ (L_{\nu}\nu)_{ERC} \simeq \Gamma^4 (L_{\nu'}'\nu')_{ERC} \sim 
\Gamma^4 ({1\over 2} N_{\gamma} \gamma) 
m_e c^2 \left\vert d\gamma\over dt'\right\vert_{ERC} \simeq $$
\be \simeq 2.4 {\sigma_T \over m_p c^2} a u_{diff} L_j \Gamma^4 
\alpha_X \gamma_{min}^{2\alpha_X} \gamma^{2(1-\alpha_X)} \, , \label{LERC} \ee
where 
\be m_e c^2 \left\vert d\gamma\over dt'\right\vert_{ERC} \simeq
(16/9)c\sigma_T u_{diff}\Gamma^2\gamma^2 \, , \label{dgdtec} \ee
\be N_{\gamma} \simeq {4 \over 3} \pi a^3  n_{\gamma}'\, , \label{Nvsn} \ee
and 
\be \nu \simeq (4/3) \Gamma^2 \gamma^2 \nu_{ext} \, . \label{nug} \ee

The external radiation field in quasars, as seen in the jet frame  at 
a distance
\be r_{fl} ={a \over \theta_j} \simeq {c t_{fl} \Gamma \over \theta_j}
\simeq 2.6 \times 10^{17} \, 
{(t_{fl}/1{\rm \, d})(\Gamma/10)^2 \over (\theta_j \Gamma)} \, {\rm cm}, 
\label{rfl} \ee
where $t_{fl}$ is the time scale of duration of flares observed in OVV quasars,
is dominated by two components, broad emission lines and 
near infrared radiation from hot dust. 
The broad emission lines provide radiation field with
$h \nu_{ext} \simeq 10$ eV and
\be u_{diff(BEL)} \simeq {L_{BEL} \over 4 \pi r_{fl}^2 c}
\sim 3.9 \times 10^{-2} {L_{BEL,45} (\theta_j \Gamma)^2 \over 
(t_{fl}/1{\rm \, d})^2 (\Gamma/10)^4} \, 
{\rm erg \, cm}^{-3} \, . \ee
The spectrum of the infrared radiation from hot dust, on the other
hand, peaks around
$h\nu_{ext} = 3 kT \sim 0.26 (T/1000)$ eV, and at a distance 
$r_{fl} < r_{IR} = \sqrt{L_d/4\pi \sigma_{SB} T^4}$ , 
has energy  density
\be u_{diff(IR)} \simeq \xi_{IR}{ 4 \sigma_{SB} \over c } T^4
\simeq 7.6 \times 10^{-3} \xi_{IR} \left( T \over 1000 {\rm \, K} \right)^4 \,
{\rm erg \, cm}^{-3} \, , \ee
where $T$ is the temperature of dust, $\sigma_{SB}$ is the 
Stefan-Boltzman constant, and $\xi_{IR}$ is the fraction of the central
source covered by the innermost parts of a 
dusty molecular torus. 
For $\Gamma \sim 10$ and typical 1 -- 20 keV spectral index 
$\alpha_X \simeq 0.6$ 
(Kii et al. 1992; Kubo et al. 1998), Comptonization of broad 
emission lines gives 
\be  (L_{\nu}\nu)_{C(BEL)} \simeq 
6.4 \times 10^{43} {L_{BEL,45} \over (t_{fl}/1{\rm \, d})}
\left( {\rm h}\nu \over 1\,{\rm \, keV} \right)^{0.4} \gamma_{min}^{1.2}
L_{j,46} (\theta_j \Gamma) \, {\rm erg \, s}^{-1} \, , \label{ECUV} \ee
while Comptonization of near infrared radiation gives
\be  (L_{\nu}\nu)_{C(IR)} \simeq
5.1 \times 10^{42}\left(\xi_{IR} \over 0.1\right) 
\left(T_{IR} \over 1000{\rm \, K}\right)^4
 \left( t_{fl} \over 1{\rm \, d}\right) 
\left({\rm h}\nu \over 1{\rm \, keV} \right)^{0.4} \gamma_{min}^{1.2}  
L_{j,46} \, {\rm erg \, s}^{-1} \, . \label{ECIR} \ee
Note that in the case of Comptonization of UV photons, radiation at 1 keV is 
produced by electrons which are only weakly relativistic, 
with $\gamma \sim 1$ (see Eq. \ref{nug}) and therefore this implies 
that $\gamma_{min}$ also needs to be $\sim 1$. In the case of
Comptonization of near infrared radiation, $\gamma_{min} \le 50/\Gamma$.

As one can see from Eqs. (\ref{ECUV}) and (\ref{ECIR}), {\sl Comptonization
of external radiation by relativistic electrons in the proton-electron jets 
gives 1 keV luminosities which are $\sim 100/L_{j,46}$ times smaller 
than observed.} 
\smallskip

Let us now determine whether the observed X--ray luminosities can be 
produced by the pure proton-electron jets via the synchrotron 
self-Compton (SSC) process, in addition to the ERC process responsible
for the hard $\gamma$--ray emission.  Luminosity of the SSC radiation can 
be estimated using the formula
$$ (L_{\nu}\nu )_{SSC} \simeq \Gamma^4  (L_{\nu'}\nu' )_{SSC} \simeq
\Gamma^4 ({1\over 2} N_{\gamma} \gamma) m_ec^2
\left\vert d\gamma\over dt'\right\vert_{SSC} \simeq $$ 
\be \simeq {2 \sigma_T \over 3 \pi m_p c^3} 
{L_{syn} L_j \over a \Gamma^2} \alpha_X
\gamma_{min}^{2\alpha_X} \gamma^{2(1-\alpha_X)} \, ,  \label{LSSC} \ee
where
\be \left\vert d \gamma \over dt' \right \vert_{SSC} = 
{4 c \sigma_T \over 3 m_e c^2} u_{syn}' \gamma^2 \, , \label{dgdt} \ee
\be u_{syn}' \simeq {L_{syn} \over 2 \pi c a^2 \Gamma^4}
\, , \label{usyn} \ee
and 
\be \nu \simeq (4/3) \gamma^2 \nu_{syn,m} \, , \label{nusyn} \ee
where $h \nu_{syn,m} \sim 0.1$ eV  is the typical location of the synchrotron
spectrum peak in OVV quasars (cf. Fossati et al. 1998). 

As it is apparent from Eq. (\ref{nusyn}), production of 1keV radiation 
by SSC process involves electrons with $\gamma \sim 100$. Therefore,  
$\gamma_{min}$ is not restricted to such low values as in the case of
the ERC processes, and, in principle, for $\gamma_{min} \sim 100$,
the SSC model can reproduce the observed soft X--ray luminosities:
\be (L_{\nu}\nu )_{SSC} \sim 7.3 \times 10^{45} 
{L_{syn,47} L_{j,46} \over (t_{fl}/1d) (\theta \Gamma)}
\left( {\rm h}\nu \over 1{\rm \, keV} \right)^{0.4} (\gamma_{min}/100)^{1.2}
\, {\rm erg \, s}^{-1} \, ,\, \label{LSSC1} \ee
where as before we used $\Gamma=10$ and $\alpha_X = 0.6$.
However, since the electrons which produce X--ray spectra via the 
SSC process have the same energy 
range as those electrons which produce $\gamma$--rays above 1 MeV,
the spectral slopes of both should be similar.  The observations show
that this is not the case; the $\gamma$--ray spectra above 1 MeV
are much steeper than the X--ray spectra in the 1 -- 20 keV range, typically
by $\Delta \alpha \simeq 0.5$ (Pohl et al. 1997).
This contradiction can be eliminated 
by assuming that most of the electrons are 
injected with $\gamma \ge 500$. 
The X--rays are then produced by electrons which reach energies appropriate
for X--ray production ($100 < \gamma < 500$ for 1 -- 25 keV) by 
radiative energy losses and the resulting slope is $\alpha_X \simeq 0.5$
(Ghisellini et al. 1998; Mukherjee et al. 1999). 

Summarizing this section we conclude that:

\refitem
$\bullet$ Production of hard X--ray spectra by ERC process requires 
$\gamma_{min} < 10$ and the pair to proton number ratio 
\be {n_{pairs} \over  n_p} \sim  50 \, {L_{SX,46} \over L_{j,46}} . \,
\label{npnp} \ee
where $L_{SX} \sim 10^{46}$ erg s$^{-1}$ is the typical luminosity 
observed in OVV quasars around 1 keV;
 
\refitem
$\bullet$ Hard X--ray spectra  can  be produced by pure
proton-electron jets via the SSC process, but this requires extremely high 
values of minimum electron injection energies. 

\section{PAIR PRODUCTION AND VARIABILITY}

We propose a scenario where jets are launched as  proton-electron 
outflows in the innermost parts of the accretion flow and are loaded 
by pairs due to interactions with hard X--rays / soft $\gamma$--rays 
produced in 
the hot accretion disc coronae.   This can well occur via a two-step 
process. The first step is Compton boosting of coronal photons (with 
initial energy of 100 -- 300 keV) up to few MeV by cold electrons in 
the outflow propagating through the central region (Begelman \& Sikora 
1987).  Provided that luminosity of the coronal radiation at $> 100$ keV
is $L_{s\gamma} \sim 10^{46}$ erg s$^{-1}$, as can be deduced from 
extrapolation of $2-10$ keV spectra observed in non-OVV radio-loud 
quasars (see, e.g., Cappi et al. 1997; Xu et al. 1999),
one can find that  opacity for the above interactions is very high,
\be \tau_{e\gamma} \simeq n_x r_{corona} \sigma_T 
\sim 15 {(L_{s\gamma}/10^{46}{\rm erg \, s}^{-1}) \over (h\nu/200 {\rm \, keV}) 
(r_{corona}/3 \times 10^{15}{\rm cm})} . \, \ee
This means that each electron in the inner parts of the outflow 
(essentially forming a ``proto-jet'') produces on the order of 10 or
more 1 -- 3 MeV photons. The second step is the absorption of MeV 
photons by the coronal (100 -- 300 keV) photons in the pair creation process.
The pairs created in such a manner are dragged by the jet, but before leaving
the compact coronal radiation field, they produce a second generation
of MeV photons, and 
they in turn produce next generation of pairs. 
Such pair cascade can continue until the time  when the proto-jet
becomes opaque for coronal  radiation, i.e., when 
$n_e r_{corona} \sigma_T \sim 1$. Within this limit, the electron/positron
flux integrated over the cross-section of the proto-jet  
can reach the value
\be \dot N_e \simeq n_e c \Omega_i r_{corona}^2 \simeq 
{c \Omega_i  r_{corona} \over \sigma_T} \, , \ee
where $\Omega_i$ is the initial solid angle of the outflow.
Comparing this electron/positron flux with the total proton flux 
\be \dot N_p \sim {L_j \over \Gamma m_p c^2} \, , \ee
we find that proton-electron winds can be loaded by pairs in the central
compact X--ray source up to the value
\be {n_{pairs} \over n_p} 
\simeq {\dot N_e/2 \over \dot N_p} \simeq 
{m_p c^3 \over 2\sigma_T} {r_{corona} \Omega_i \Gamma \over L_j} 
\simeq 30 \, {r_{corona} \over 3 \times 10^{15}{\rm cm}} \,
{\Omega_i\over L_{j,46}} \, {\Gamma \over 3} \, . \ee
This corresponds roughly to the pair content given by Eq. (\ref{npnp}), 
provided that $\Omega_i$ is not very small. Note that a large initial 
opening angle of the central outflow is expected, as jets with 
$L_j \sim 10^{46}$ erg s$^{-1}$ carry too much momentum to be effectively
collimated by the innermost parts of the accretion disc corona.
Due to radial quasi-expansion, the outflow is rapidly diluted and can 
be collimated to the narrow jet by disc winds at $r > 100 $ gravitational 
radii (cf. Begelman 1995).  

It should be mentioned here that loading of quasar jets by pairs via 
absorption of $\gamma$--rays produced within the jet 
by external radiation field
has been also proposed by Blandford and Levinson (1995) (BL95).  However, 
their scenario is very different from ours in many respects. 
In the BL95 model, both pairs and nonthermal radiation are produced over 
several decades of distance; in our model, pair production is taking 
place at the base of a jet, whereas nonthermal radiation is 
produced at distances $10^{17}$ -- $10^{18}$ cm.
 The BL95 model involves relativistic electrons, and  pairs are produced 
by absorption of nonthermal radiation extending up to GeV 
energies, while our pair loading  scenario involves cold electrons 
(as measured in the jet comoving frame), and pairs are produced 
by absorption of photons with energy 1 -- 3 MeV. In their scenario,
Comptonization of the UV bump by relativistic pair cascades leads to 
a production of a  power-law X--ray spectrum which is softer than that
observed in OVV quasars; in our scenario --- in the region where pairs 
are injected, i.e., at the base of a jet --- the UV bump is
Comptonized only by cold pairs, and this leads to a production of 
radiation only around $h \nu_{UV} (\Gamma/3)^2 \sim 100$ eV. 
Due to the wide opening angle of a jet at its base, this radiation 
is much less collimated than nonthermal radiation produced at larger 
distances, and therefore in OVV quasars, it may be relatively
inconspicuous. The $100$ eV excess can be detectable eventually in steep 
spectrum radio loud quasars, which have jets pointing further away from our 
line of sight. However, due to absorption by the ISM in the host, or our own 
Galaxy, this excess is predicted to be weak, and difficult to detect.  

Another attractive feature of our model is that the pair loading of 
a jet via interactions of the proto-jet with the 
hard X--rays / soft $\gamma$--rays produced by accretion disk coronae
can be responsible for fast ($\sim$ day) 
variability observed in OVV quasars. We know from observations that the 
(presumably isotropic) X--ray emission from Seyfert galaxies -- 
and thus, by analogy, the non-jet, isotropic component in radio loud quasars 
-- is rapidly variable (although in OVV quasars, this component is  
``swamped'' by stronger, relativistically boosted flux).  This suggests that 
the corona is likely to have dynamical character:  it may be powered by 
magnetic flares, or else by a possible instability of the innermost region of
the accretion disk.  In either case, the jet is expected to be loaded by 
pairs non-uniformly and non-axisymmetrically.  The patches of the local 
pair excesses in a jet suffer large radiation drag (Sikora et al. 1996) 
and are forced to move slower than the surrounding gas.
Therefore, they  provide natural sites for shock formation and
particle acceleration.

While the above mechanism is viable as an explanation of rapid X--ray
and $\gamma$--ray variability observed in OVV quasars, we note that pair 
density variations, as modulated by magnetic flares in the disk, 
are too rapid to produce variability of the {\sl radio} flux.  The long-term
(months to years) variability in OVV quasars observed in all spectral
bands including radio, are more likely to result from modulation of
the variable flux of protons. Such modulation can be induced by the 
variability of the accretion rate in the inner parts of the accretion disc.
The observed long term optical  variability in both radio-loud and radio-quiet
quasars supports this view (see, e.g., Giveon et al. 1999 and
references therein).

\section{SUMMARY}

\refitem$\bullet$ Models of quasar jets  consisting purely of 
e$^+$e$^-$ pairs can be excluded because they predict much larger 
soft X--ray luminosities than observed in OVV quasars.
On the other hand, models with jets consisting solely of proton-electron 
plasma jets are excluded, as they predict much weaker nonthermal 
X--ray radiation than observed in OVV quasars. Spectra of nonthermal flares in 
those objects {\sl can} be explained in terms of a simple homogeneous 
ERC model, provided that the number of pairs per proton reaches values 
$\sim 50\, (L_{SX}/L_j)$.

\refitem $\bullet$ We suggest that initially, jets consist mainly of 
proton-electron plasma (where the protons provide the inertia to
account for the kinetic luminosity of the jet), and subsequently are 
loaded by e$^+$e$^-$ pairs by interactions with hard X--rays / soft 
$\gamma$--rays from hot accretion disc coronae. This requires 
that the coronal temperatures reach values $\sim 100$ keV, which 
are consistent with observations of Seyfert galaxies, with spectra 
uncontaminated by relativistic jets.  

\refitem $\bullet$ Non-steady and non-axisymmetric pair loading of 
jets by X--rays from
magnetic flares in the corona can be responsible for short term 
($\sim$ day) variability observed in OVV quasars. It should be emphasized
here that alternative mechanisms of variability, such as modulation of 
the total energy flux in a jet by accretion rate or precession of a jet 
(cf. Gopal-Krishna \& Wiita 1992) cannot operate on such short time scales.
The lack of rapid (time scale of $\sim$ day), high amplitude 
variability in the UV band of radio lobe dominated quasars supports 
this view. 

\refitem $\bullet$ The dissipative sites in quasar jets, where 
electrons/positrons are accelerated and produce nonthermal flares observed 
in OVV quasars, can be provided by shocks
produced by collisions between inhomogeneities induced by non-uniform
pair loading of the proto-jets. These shocks (and therefore particle 
acceleration) can be amplified eventually at a distance 0.1 -- 1 pc due 
to reconfinement of a jet by the external gas pressure (Komissarov \& 
Falle 1997; Nartallo et al. 1998). 

\bigskip

We are grateful to Annalisa Celotti, the referee, whose thoughtful comments
helped us to substantially improve the paper. 
M.S. thanks Annalisa Celotti and Paolo Coppi for stimulating  discussions,
and, in particular, for pointing 
out that the pair-loading process in proto-jets can saturate.  
We would like to thank the Institute for
Theoretical Physics at U.C. Santa Barbara for its hospitality.
This project was partially supported by 
ITP/NSF grant PHY94-07194, NASA grants and contracts to University of
Maryland and USRA, and the Polish KBN grant 2P03D 00415.

\end{document}